\documentclass[10pt,journal,doublecolumn]{IEEEtran}

\usepackage[square,sort,comma,numbers]{natbib}
\usepackage{amsfonts,epsfig,graphicx,tabularx,multirow,fancyhdr,subfigure}
\usepackage{multirow} % For tables
\pdfoutput=1
\usepackage{mathtools}

\usepackage{array}
\usepackage{mdwmath}
\usepackage{lineno}

\begin{document}

\title{\raggedright{Electrocardiogram Signal Denoising using Nonlocal Wavelet Transform Domain Filtering}}
\author{\raggedright{Santosh~Kumar~Yadav, Rohit~Sinha, Prabin~Kumar~Bora \\% <-this % stops a space
Department of Electronics and Electrical Engineering,\\ Indian Institute of Technology Guwahati, Guwahati-781039, Assam, India \\
E-mail: \{y.santosh, rsinha, prabin\}@iitg.ernet.in
}}

\maketitle
%\linenumbers

%\textbf{Abstract:} ECG signals are usually corrupted by baseline wander, power-line interference, muscle noise, etc. and numerous methods have been proposed to remove these noises. However, in case of wireless recording of the ECG signal it gets corrupted by the additive white Gaussian noise (AWGN). For the correct diagnosis, removal of AWGN from ECG signals becomes necessary as it affects the all the diagnostic features. The natural signals exhibit correlation among their samples and this property has been exploited in various signal restoration tasks. Motivated by that, in this work we propose a nonlocal wavelet transform domain ECG signal denoising method which exploits the correlations among both local and nonlocal samples of the signal. In the proposed method, the similar blocks of the samples are grouped in a matrix and then denoising is achieved by the shrinkage of its two-dimensional discrete wavelet transform coefficients. The experiments performed on a number of ECG signals show significant quantitative and qualitative improvement in denoising performance over the existing ECG signal denoising methods.


%\IEEEdisplaynontitleabstractindextext
%\IEEEpeerreviewmaketitle

\section{Introduction}

 \IEEEPARstart{E}{CG} signal is one of the most popular diagnostic means which provides an electrical picture of the heart and information about different pathological conditions. These signals originate from the heart and pass through the tissues with different characteristics to reach up to the several recording leads placed on the skin of the subject. Because of the path deformities and external electrical disturbances, ECG signals become noisy. The typical noises in the ECG include baseline wander, power-line interference, muscle noise, etc. In recent years, the biotelemetry has become a dominant means of monitoring the cardiac condition of ambulatory patients~\cite{Jing-Jhang1999,Alesanco-Garcia2010}. Also, to detect arrhythmias and cardiac abnormalities, wireless ambulatory ECG recording is now routinely used~\cite{RafiAhmed}. In such cases, the ECG data is sent through the channel (wireless, telephone lines, etc.) to a remote location where it is analyzed. In this process, it gets corrupted by the channel noise. For the correct diagnosis, the removal of noise from such ECG signals becomes necessary.

 In the past, a number of algorithms have been proposed for ECG signal denoising ~\cite{huang1998empirical,Agante1999,RSemani2007,OSayadi2008,BlancoVelasco2008,Gao-Sultan-Hu-Tung2010}. Among those, the methods based on the discrete wavelet transform (DWT) coefficient shrinkage \cite{Kestler1998,Agante1999,Gao-Sultan-Hu-Tung2010} and the empirical mode decomposition (EMD) \cite{huang1998empirical,BlancoVelasco2008} have emerged as two popular groups. In the former group, a good estimate of clean ECG signal is obtained by discarding the lower magnitude DWT coefficients followed by the inverse wavelet transform. In the latter group, the estimate of clean ECG signal is obtained by discarding first few intrinsic mode functions (IMFs) since these account for the high frequency variations present in the signal. However, this process is reported to distorts the QRS complexes. In~\cite{Weng2006} the portions of the first few IMFs those correspond to the QRS complexes are preserved by means of a \emph{Tukey} window. In~\cite{Kabir2012481} a hybrid EMD-wavelet method that combines the windowed EMD with wavelet soft-thresholding has been proposed to further improve the denoising performance.

 The nonlocal means (NLM) method~\cite{Buades2005} is a very successful image denoising method. Recently, it has been applied for ECG signal denoising~\cite{Tracey-Miller2012} and is shown to outperform the hybrid EMD-wavelet method for a number of ECG signals. The NLM method was originally developed for image denoising with the assumption that the underlying clean image has several pixels with similar neighborhood. In normal cases, the ECG signals are almost structurally repetitive and thus possess such redundancy. In one-dimensional NLM denoising proposed in~\cite{Tracey-Miller2012}, the estimates of the underlying clean signal samples are obtained by weighted averaging of the samples having similar neighborhoods. The applied weighting is proportional to the similarity in the neighborhood and is independent of the temporal location of the samples. As a result, the samples with quite similar neighborhoods are given higher weights whereas lower weights are assigned to the samples with dissimilar neighborhoods. Thus, it directly exploits the nonlocal similarity present in the signal.

 The NLM algorithm uses a sample-based approach in which each sample is estimated independently. In other words, the estimate of a sample at one location does not contribute to the estimation of other samples even if those are in close proximity. The nonlinear filtering methods like the shrinkage of the DWT coefficients do not face such drawback as these rely on the inherent sparsity of the clean signal in the transform domain. However, the DWT shrinkage based methods could not exploit the nonlocal redundancy present in the signal. On combining the transform based approach and the block-based NLM approach, their relative advantages can be exploited. The similar idea has already been explored for image denoising~\cite{BM3D2007,Priyam2009,Rajwade-Banerjee} but is yet to be explored for the biomedical signals like ECG, EEG, etc. In this paper we propose a novel ECG denoising method which exploits local as well as nonlocal similarity in the signal. In the proposed method, the similar blocks of samples are estimated in a collaborative manner. The denoising is accomplished by the shrinkage of the two-dimensional (2D) DWT coefficients of the matrix formed with these similar blocks. This process is repeated for each of the overlapping blocks resulting in several estimates for a sample. The final estimate is found by averaging these estimates.

The remainder of this paper is organized as follows. In Section~\ref{sec:existingmethods}, the denoising problem is formulated and the existing NLM denoising method for ECG signals is described. The details of the proposed algorithm along with the parameter tuning are given in Section~\ref{sec:proposedmethods}. The experimental results and discussions are presented in Section~\ref{sec:results} and Section~\ref{sec:Discussion}, respectively. Section~\ref{sec:conclusion} concludes the paper.

%%%%%%%%%%%%%%%%%%%%%%%%%%%%%%%%%%%%%%%%%%%%%%%%%%%%%%%%%%%%%%%%%%%%%%%

\section{Prior Art}
\label{sec:existingmethods}
\subsection{Problem Formulation}
 Denoising, as an inverse problem, addresses the recovery of a clean signal from its noisy observation given the knowledge about the parameters of the noise. The additive white Gaussian noise (AWGN) is commonly used to model the channel noise. Assuming the ECG signal contaminated by AWGN, the observed ECG signal is given by,

\begin{equation}
  v[i]=u[i]+\eta[i],~~~i=1,2,\ldots,N
\end{equation}
 where $u[i]$ is the clean signal and $\eta(i)$ is the zero mean AWGN with variance $\sigma^2$. Natural signals including ECG signals exhibit correlation among their samples and this permits their sparse representation in a suitable domain. AWGN samples are uncorrelated and do not yield a sparse representation in that domain. Hence, on using the sparsifying transforms noisy signals can be denoised. Also, the higher is the sparsification of a signal, the better is the removal of noise in the sparsifying transform domain.

\subsection{NLM Algorithm}
\label{sec:NLM}
 The NLM algorithm~\cite{Buades2005} has been one of the most popular image denoising methods in computer vision. In this method, the estimate of $i^{th}$ sample is found by weighted averaging of the samples with similar local neighborhood within a search window $N(i)$. As defined in~\cite{Tracey-Miller2012}, the clean signal is estimated as

\begin{equation}\hat{u}[i] = \frac{1}{C(i)}\sum_{j\in{N(i)}}w(i, j)v[j],\end{equation}
 where $w(i, j)$ is the weight of $j^{th}$ neighboring sample with respect to $i^{th}$ sample and \(C(i) = \sum_{j}w(i, j)\). Motivated by \cite{Van-Kocher2009}, in \cite{Tracey-Miller2012} the weight $w(i, j)$ is computed as

\begin{align} \label{eq:weighteqnNLM}
\nonumber
w(i, j) &= \exp\bigg(-\frac{\sum_{k\in{\Delta}}(v[i+k]-v[j+k])^2}{2L_{\Delta}\mu^2}\bigg) \\
&= \exp\bigg(-\frac{d^{2}(v[i],v[j])}{2L_{\Delta}\mu^2}\bigg),
\end{align}
 where $\mu$ is the bandwidth parameter, $\Delta$ represents the neighborhood containing $L_{\Delta}$ samples surrounding the sample of interest. The term $d^{2}(\cdot,\cdot)$ represents the squared Euclidean distance between the set of chosen samples accounting for the similarity of their neighborhoods. The same procedure is repeated for each sample and finally we are left with the denoised signal. In \cite{Buades2005}, the similarity measure was taken as the Euclidean distance between Gaussian neighborhood. However, on experimentation such weighting of neighborhoods is not found to be effective in the case of ECG signals.

 In the NLM algorithm, the samples of the clean ECG signal with similar neighborhoods are assumed to be quite close in value. Thus on averaging those samples, the signal can be denoised if the considered noise is zero-mean. Such an assumption may be reasonable for the digital images but it may not hold for the QRS complexes in ECG signals. Further, the similar samples having dissimilar neighborhoods are often given lower weights. On the other hand, the highly noisy samples having quite similar neighborhoods are given higher weights. Both of these facts lead to the noisy estimate of a sample. Also in the NLM algorithm, the estimate of a sample does not contribute to the estimation of even the nearby samples. In the next section, we discuss that how the proposed method addresses these problems.

%%%%%%%%%%%%%%%%%%%%%%%%%%%%%%%%%%%%%%%%%%%%%%%%%%%%%%%%%%%%%%%%%%%%%%%%%%%%%%%%%%%%%%%%%%%%%%%%%%%
\section{NLWT Denoising Method}
\label{sec:proposedmethods}

 The proposed nonlocal wavelet transform (NLWT) domain algorithm differs from the existing NLM algorithm in two aspects: the block-based processing and the transform domain collaborative filtering. In the block-based processing, we estimate a block of samples instead of individual samples. As the successive blocks are kept overlapping, multiple estimates are produced for a sample. These estimates are then averaged to find the final estimate. This averaging adds an extra layer of smoothing which helps in addressing the problem of inappropriate weights assigned to the samples as mentioned in the context of the NLM algorithm. Next, while doing the weighted averaging of the blocks for denoising, the intra-block correlations are not exploited. To do so, the shrinkage of 2D orthogonal transform coefficients of a matrix having similar blocks as its columns can be used. The shrinkage of the 2D DWT coefficients has been one of the most simple yet effective denoising methods in image processing as it exploits inter-row and inter-column correlations. Hence, we have used the 2D DWT as an orthogonal transform in the proposed method. The proposed algorithm is described in the following subsections.

\begin{figure*}[ht]
\centering
\subfigure{
\includegraphics[scale=0.8]{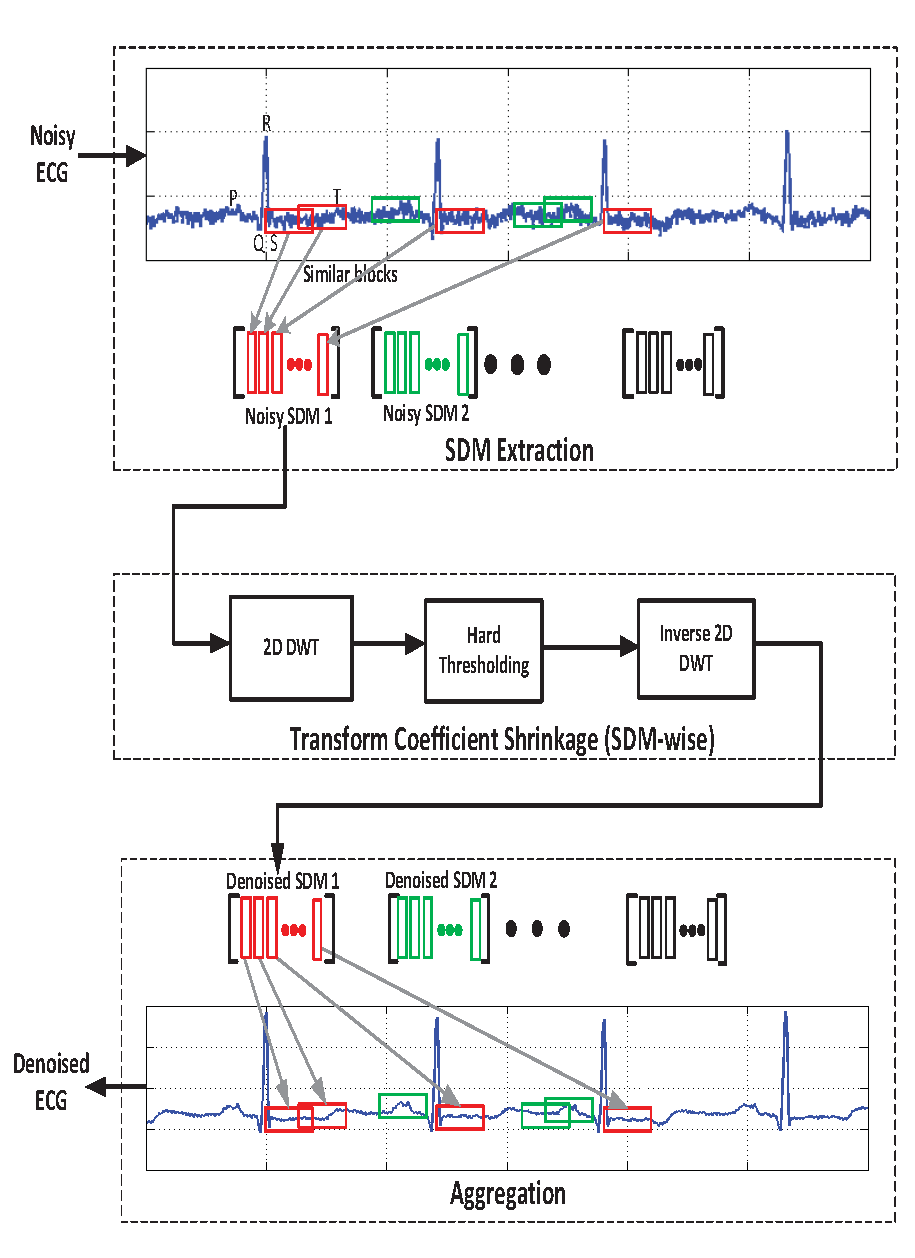}}
\caption{Block diagram of the proposed NLWT method. The blocks of samples that are marked in same colors are similar and grouped in an SDM. The transform coefficient shrinkage is applied on each SDM. In the aggregation step, the estimate for the overlapped portions are obtained by the weighted averaging of the involved sample estimates.}
\label{fig:ecg1}
\end{figure*}

 \subsection{Notations}
  For mathematical description of the proposed method, we introduce some notations. Let $D_i$ be an operator that extracts a vector corresponding to the block of samples centered at $i^{th}$ location from the signal vector $\textbf{v}$. Hence, $D_{i}\textbf{v} = [v[i-L]\ldots v[i]\ldots v[i+L] ]^{T}$ is a vector of dimension $2L+1$ corresponding to the $i^{th}$ reference block. Clearly, $D_{j}\textbf{v}$ with $j\in\{i-M,..., i-1, i, i+1,..., i+M\}$ is the neighboring blocks to be compared with the $i^{th} $reference vector with a similarity measure $l(D_{i}\textbf{v},D_{j}\textbf{v})$ over a search window of size $2M+1$. We define the set of locations of similar blocks corresponding to $i^{th}$ reference block as $\Omega_i=\{j~|~ l(D_{i}\textbf{v},D_{j}\textbf{v})\leq \tau\}$, where $\tau$ is a predetermined matching threshold. Let $\mathfrak{D}_{\Omega_i}$ be a matrix containing these similar blocks as its columns extracted from the locations $\Omega_i$ and termed as the \emph{similarity data matrix (SDM)}~\cite{Zhang-Feng-Wang2013}. The sorting of the columns of the SDM $\mathfrak{D}_{\Omega_i}$ in descending order of similarity with the reference block can also be incorporated. It brings the reference block in the first column for notational simplicity. Specifically, $\mathfrak{D}_{\Omega_i}=[D_{i}\textbf{v}~ D_{\Omega_i(1)}\textbf{v}~D_{\Omega_i(2)}\textbf{v}~\cdot\cdot\cdot~D_{\Omega_i(m-1)}\textbf{v}]$ is a $(2L+1) \times m$ matrix with $m$ columns similar to the $i^{th}$ reference block.

 \subsection{Proposed NLWT Algorithm}
 The proposed algorithm involves three sequential steps: the \emph{SDM extraction}, the \emph{transform coefficients shrinkage}, and the \emph{aggregation}. The overall block diagram of the proposed NLWT method is shown in Fig.~\ref{fig:ecg}. In the first step, the SDMs corresponding to each of the reference blocks are formed by searching similar blocks. In the second step, each of the SDMs is denoised by the shrinkage of its wavelet transform coefficients. In the last step, the blocks from these denoised SDMs are then returned to their original location to get the denoised signal. The further details of these steps are discussed below.

 \subsubsection{SDM extraction} The SDM extraction step starts with selecting a reference block and searching for similar blocks to the reference block within a search window. At the maximum, $m$ most similar blocks including the reference block are allowed in an SDM. The next reference block is selected by shifting $k$ samples and the above process is repeated to form the next SDM. The number of SDMs formed for a given ECG signal depends on the amount of overlap kept between successive reference blocks and is parameterized by the shift parameter $k$. The range of the shift parameter is kept as $0<k<(2L+1)$ which ensures at least one sample overlap between successive reference blocks. For an ECG signal length of $N$ samples, the total number of SDMs turns out to be $1+\lfloor\frac{N-(2L+1)}{k}\rfloor$, where $\lfloor \rfloor$ denotes the floor function.

\begin{figure}[!ht]
\centering
\subfigure{
\includegraphics[scale=0.38]{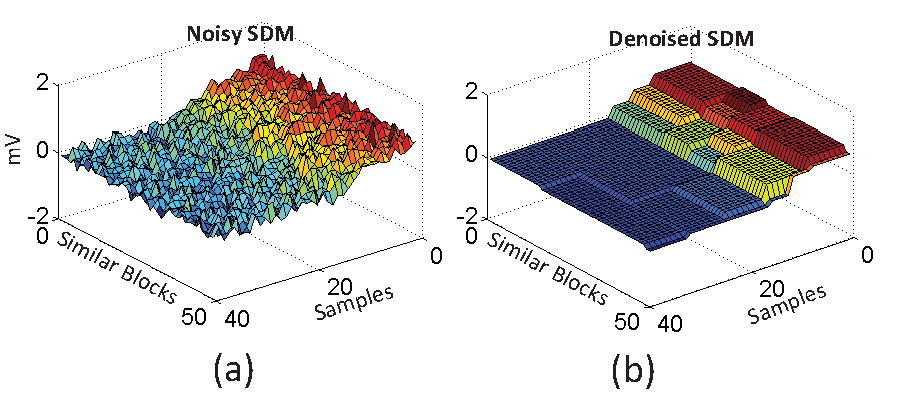}}
\caption{Illustration of a SDM extracted from noisy version of ECG signal number 100 and corresponding denoised SDM by the shrinkage of its 2D DWT coefficients.}
\label{fig:ecg}
\end{figure}

 The purpose of this step is to find the blocks whose underlying noise-free counterparts are similar. As we do not have the direct access to the clean signal, a lighter smoothing is done before finding the similarity. Moreover, it would be better to measure the similarity in a locally learned feature space as it can adapt to the structural changes within the blocks and also reduces the complexity of distance computation. For the same, the principal component analysis (PCA) is performed with the blocks  $D_{j}\textbf{v}$,  where $j\in\{i-M,..., i-1, i, i+1,..., i+M\}$. A projection matrix $A_{i}$ is formed by choosing only a few principal components. The similarity between $D_{i}\textbf{v}$ and $D_{j}\textbf{v}$ is then measured as,

\begin{equation}
\label{eq:sim_mes1}
l(D_{i}\textbf{v},D_{j}\textbf{v}) = ||A_{i}(D_{i}\textbf{v})-A_{i}(D_{j}\textbf{v})||_{2}^{2},
\end{equation}
where $A_{i}(D_{j}\textbf{v})$ is the projection of $j^{th}$ block on the feature space. If $l(D_{i}\textbf{v},D_{j}\textbf{v})\leq \tau$, the block $D_{j}\textbf{v}$ is selected as being similar to $D_{i}\textbf{v}$ and its location is incorporated in the set $\Omega_i$. Finding out the principal components to form SDMs is computationally intensive task, so one can use the discrete cosine transform (DCT) as a compromise. The SDM extraction process can be better understood from the Fig.~\ref{fig:ecg1}. Fig.~\ref{fig:ecg} (a) shows an example of SDM extracted from a noisy ECG signal. This highlights the structural similarity among the blocks captured in an SDM.

\subsubsection{Transform coefficient shrinkage} The SDM $\mathfrak{D}_{\Omega_i}$ being a collection of similar blocks, has a homogeneous structure. This redundancy can be exploited by a suitable sparsifying transform. In this work we have used 2D DWT for this purpose. Each of the noisy SDMs is transformed into the wavelet domain and then transform coefficients are thresholded. Corresponding denoised SDMs are then obtained by taking inverse transform. The entire procedure can be mathematically expressed as,

\begin{equation}
\label{eq:CoreDen1}
\widehat{\mathfrak{D}}_{{\Omega_i}}= \mathcal{W}^{-1}(\mathcal{S}(\mathcal{W}(\mathfrak{D}_{\Omega_i}), \lambda_{i}))~~~~\forall~i,
\end{equation}
where $\mathcal{W}$ is the 2D DWT operator, $\mathcal{S}(\cdot, \lambda_{i})$ is hard-thresholding operator given as,

\begin{eqnarray}
\label{eq:CoreDen_Thresh}
  \mathcal{S}(t, \lambda_{i})  = \left\{\def\arraystretch{1.2}%
  \begin{array}{@{}c@{\quad}l@{}}
    t & \text{if}~~~|t|\geq \lambda_i\\
    0 & \text{otherwise}\\
  \end{array}\right.
\end{eqnarray}
and $\lambda_{i}$ is the shrinkage threshold corresponding to $\mathfrak{D}_{\Omega_i}$.

 The shrinkage of the 2D DWT transform coefficients of the noisy SDMs exploits both inter-row (local) similarity and inter-column (nonlocal) similarity. On comparing Fig.~\ref{fig:ecg} (a) and Fig.~\ref{fig:ecg} (b), the structural similarity of the SDM is well preserved with the proposed 2D DWT coefficient shrinkage.

\subsubsection{Aggregation} This is the final step in which the estimated blocks are sent to the temporal locations where they were extracted from (see Fig.~\ref{fig:ecg1}). Note that, a sample would be present in a number of blocks from one or more SDMs resulting in its multiple estimates. The weighted average of these multiple estimates is taken as the final estimate of that sample. The natural choice for the weight of aggregation is the inverse of the variance of the denoised SDM  $\widehat{\mathfrak{D}}_{{\Omega_i}}$. The intuition behind this choice is that the estimated blocks with the sharp gradient are to be given lesser importance than with the smooth ones. Assuming the independence among the estimated columns of $\widehat{\mathfrak{D}}_{{\Omega_i}}$, the weight $\omega_i$ can be given as
\begin{eqnarray}
\label{eq:CoreDen2}
\mathrm{\omega}_i =  1/(\mathrm{\textit{N}}_{\mathcal{S},i}\,\sigma^2),
\end{eqnarray}
where $\mathrm{\textit{N}}_{\mathcal{S},i}$ is the number of non-zero coefficients in $\mathcal{S}(\mathcal{W}(\mathfrak{D}_{\Omega_i}), \lambda_{i})$ and $\sigma^2$ is the variance of the noise. Thus, the estimated signal can be given as

\begin{equation}
\label{eq:CoreDen3}
\widehat{u}[i] = \frac{\sum_{j=1}^{n}\omega_{j}\,\widehat{\mathfrak{D}}_{\Omega_j}[\mathfrak{L}_{i}] }{\sum_{j=1}^{n}\omega_{j}} ~~~\forall~i,
\end{equation}
where $n$ is the total number of estimates and $\mathfrak{L}_{i}$ is $n$-index pointer for the locations of $i^{th}$ sample in/across the SDMs.

 This weighting has the intuitive appeal of avoiding smoothing the QRS complex, as the priority is given to the homogeneous blocks which have undergone through aggressive thresholding. In such SDMs, the number of coefficients retained after thresholding is smaller, and hence assigned larger weights.

 \subsection{Parameters}
 \label{sec:parameterselection}
 The different parameters used in the proposed algorithm are: the block half-width, the search window half-width, the maximum number of blocks allowed in an SDM, the matching threshold, the shrinkage threshold, and the type of mother wavelets. The significance of each of these parameters is discussed below.

\begin{table*}[!ht]
\caption{Different parameters used in the proposed NLWT algorithm, their tuning ranges, and the optimal values.}
\begin{center}
\renewcommand{\arraystretch}{1.2}
\begin{tabular}{|l|c|c|c|}
\hline
    \textbf{Parameters} & \textbf{Notation} & \textbf{Range for tuning} & \textbf{Optimal value}  \\ \hline \hline
    Block half-width    & $ L $& $0.01 f_s-0.1 f_s$ & 10  \\ \hline
    Search window half-width      & $ M $& 3-5 Heart beats& 1000  \\ \hline
    Maximum number of blocks in SDM  & $m$ & -  & $2(2L+1)$ \\ \hline
    Matching threshold  & $\tau$& $1-5$~\% ~\text{of}~ $2(2L+1)$  & 1.2 \\ \hline
    Shrinkage threshold coefficient  & $c$ & $\pm 25$~\% ~\text{of}~$2\sqrt{\log (2L+1)}$ & 3.8  \\ \hline
    Mother wavelet      & - & db 1-45, sym 2-20, coif 1-5 & Haar (db2) \\ \hline
\end{tabular}
    \label{table-tab1}
    \end{center}
    \end{table*}

 \subsubsection{Block size}
 The block size denotes the number of samples in a block and is parameterized by the block half-width $L$. It should be large enough to capture the feature of interest in the signal. However, a larger value of $L$ would create bigger blocks with reduced chances of match being found within a search window and a smaller $L$  would result in a large number of blocks which in turn lead to erroneous grouping of the blocks as well as increased overall complexity. We believe that the QRS complexes as a whole have lesser chance of finding a good match in the signal than their smaller portions. Usually, the duration of the QRS complexes, in normal sinus rhythms, are in between $0.06-0.1$ second. Based on this, for a signal sampled at a rate of $f_s$ Hz, the appropriate block size is searched within the range of $0.01 f_s-0.1 f_s$. Also, the shift parameter $k$ depends on the block size. For a trade-off between performance and complexity, the overlap between successive reference blocks is kept as $50$~\% in this work, i.e., $k = L$.

 \subsubsection{Search window}
 The most trivial way is to search for the similar blocks over the entire length of signal but to reduce the complexity it is restricted within a window of length ($2M+1$), where $M$ denotes the half-width of the search window. For a smaller value of $M$ only the local-search is done while a larger value of $M$ leads to the increased complexity. Hence, a moderate value of $M$ is chosen so as to include multiple heartbeats allowing multiple QRS complexes, P-waves, and T-waves with potentially similar characteristics.

 \subsubsection{Matching threshold}
 The matching threshold is one of the most sensitive parameters of the proposed method as it controls the number of similar columns in an SDM and is denoted by  $\tau$. It should be properly selected so that several blocks with similar features can be incorporated in an SDM.  For a smaller values of $\tau$ no similarity for a reference block is found in the signal, leading to a single column SDM which makes the whole effort futile. A larger value of $\tau$ allows the dissimilar blocks being grouped in the SDMs and would lead to smoothing of the diagnostic features. For the ECG signal normalized to $\pm1$ mV range distance between two blocks of size $2L+1$ can not exceed $2(2L+1)$. Also, as this parameter directly affects the maximum number of blocks in an SDM $m$, a good selection of $\tau$ should result in a square SDM. In the proposed method, the value of this parameter is determined experimentally.

 \subsubsection{Shrinkage threshold}
 The shrinkage threshold $\lambda_i$ is a parameter which directly affects the denoising performance. It should be proportional to the noise standard deviation $\sigma$ so that the amount of shrinkage can be selected according to the noise present in the signal. We choose $\lambda_i=c_i \sigma$, where constant $c_i$ controls the smoothing. This constant can be obtained by the \emph{VisuShrink} soft-thresholding formula~\cite{Donoho1995,DONOHOjohnsten} given as $c_i=\sqrt{2\log N_i}$, where $N_i$ is the total number of transform coefficients in the $i^{th}$ SDM.

%%%%%%%%%%%%%%%%%%%%%%%%%%%%%%%%%%%%%%%%%%%%%%%%%%%%%%%%%%%%%%%%%%%%%%%%%%%%%%%%%%%%%%%%%%%%

\section{Experiments and Results}
\label{sec:results}

 The performance of the proposed method is evaluated on the MIT-BIH arrhythmia database~\cite{MoodyMIT-BIHArrhy} and the PTB diagnostic database~\cite{PTBdataset} taken from the Physionet~\cite{PhysionetGlobal} data bank. The former database is used to compare the performance with the existing techniques while the latter one is used to verify the performance of the proposed method on ECG signals with known pathology. Signal denoising methods are generally evaluated based on the ability that how close the denoised signal is to the original signal. To be consistent with the works in~\cite{Kabir2012481} and~\cite{Tracey-Miller2012}, the following performance measures is chosen for the quantitative assessment.

\textbf{Signal-to-noise-ratio improvement} ($\text{SNR}_{\text{imp}}$):
\begin{equation}
 \text{SNR}_{\text{imp}} = 10\log{\frac{\sum_{i=1}^{N}(v[i]-u[i])^2}{\sum_{i=1}^{N}(\hat{u}[i]-u[i])^2}}
 \label{eqn:SNRimp}
\end{equation}

\textbf{Mean square error} (MSE):
\begin{equation}
 \text{MSE} = \frac{1}{N}\sum_{i=1}^{N}(\hat{u}[i]-u[i])^2
 \label{eqn:MSE}
\end{equation}

\textbf{Percent root mean square difference} (PRD):
\begin{equation}
  \text{PRD} = 100 \sqrt{\frac{\text{MSE}}{\frac{1}{N}\sum_{i=1}^{N}u^2[i]}}
  \label{eqn:PRD}
\end{equation}
 In the equations (\ref{eqn:SNRimp}), (\ref{eqn:MSE}), and (\ref{eqn:PRD}) $N$ is the total number of samples in the signal under consideration and other symbols have their usual meanings as described earlier. Note that a better denoising method should have a higher $\text{SNR}_{\text{imp}}$, a lower MSE, and a lower PRD.

\subsection{Evaluation on MIT-BIH Physionet database}

 The signals in the MIT-BIH Physionet database have the sampling frequency about 360 Hz. To evaluate the denoising methods, we perform experiments with noisy ECG signals those are artificially generated by adding white Gaussian noise (AWGN) of chosen signal-to-noise ratio (SNR) levels to the raw signals taken from the database. These noisy ECG signals are then denoised by the proposed and a few recent methods. To compare the results with the methods in~\cite{Kabir2012481} and~\cite{Tracey-Miller2012}, the experimental performance using the signals numbered as 100, 103, 104, 105, 106, 115, and 215 is reported.

 The different parameters involved in the proposed method are tuned on the signal number 100 (normal sinus rhythm) of the data set. Tuning is done by varying the parameters within the suggested range given in third column of Table~\ref{table-tab1} to maximize the SNR improvement. The optimal value of parameter $L$ is found to be 10. For a given $L$, $M$ is varied from 500 to 5000 samples and $M$ = 1000 is found to be the best choice. To reduce the computational complexity we have used a fixed value of hard-threshold multiplier $c$ for all SDMs in this work. The average number of elements (samples) in an SDM is approximated to be $N_i\simeq(2L+1)^2$. Hence, the value of coefficient $c$ is experimentally selected by varying it about $2\sqrt{\log (2L+1)}$  and is found to be 3.8. Despite the fixed value of hard-thresholding parameter $c$, our approach is adaptive as we have used adaptive weighting in the aggregation step. For choosing the appropriate mother wavelet, the experiments were performed with the different wavelet families such as Daubechies (db), Symlets (sym), and Coiflets (coif). On an average, the Haar wavelet (db 2) is found to result in the best averaged performance. Similarly, the block matching threshold $\tau$ is found to be 1.2 while using five largest principal components in projection matrix. We observed that, on an average, the parameters selected as above give the minimum MSE and the minimum PRD for the chosen data set.

%%%%%%%%%%Performance on different signals at 20dB  %%%%%%%%%%%%
\begin{figure}[!ht]
\centering
\subfigure{
\includegraphics[width=8.5cm,height=8cm]{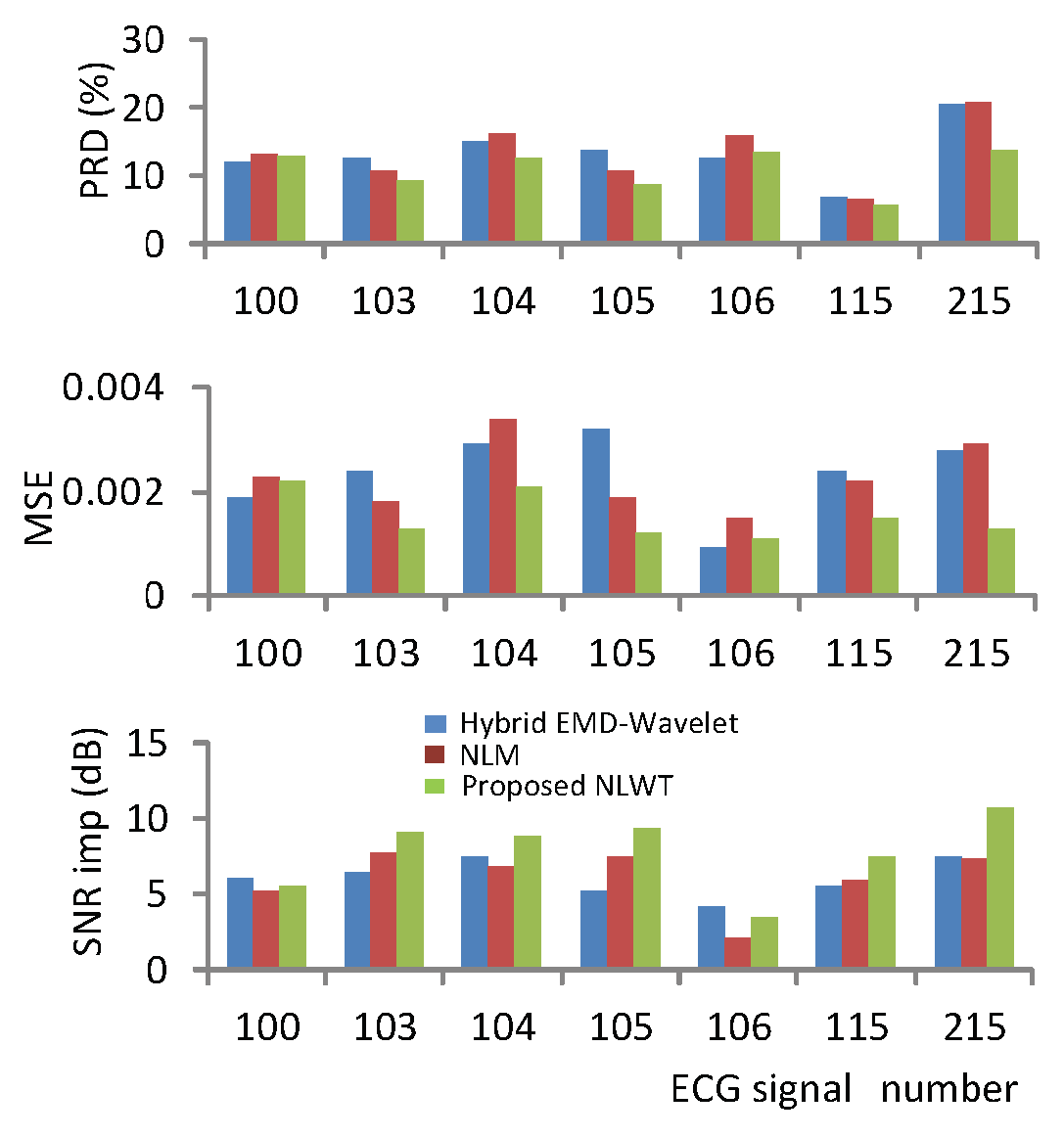}}
\caption{Comparision of the denoising performance, in terms of different measures, of the proposed and the existing methods on different ECG signals at 20 dB noise level.}
\label{fig:perfecg2}
\end{figure}

%%%%%%%%%%%%%%%%%%%%%%%%Average performance%%%%%%%%%%%%%%%%%%%%%%%%%%%%%%%
\begin{figure}[!ht]
\begin{center}
\advance\leftskip-1.2cm
\subfigure{
\includegraphics[width=8.3cm,height=4cm]{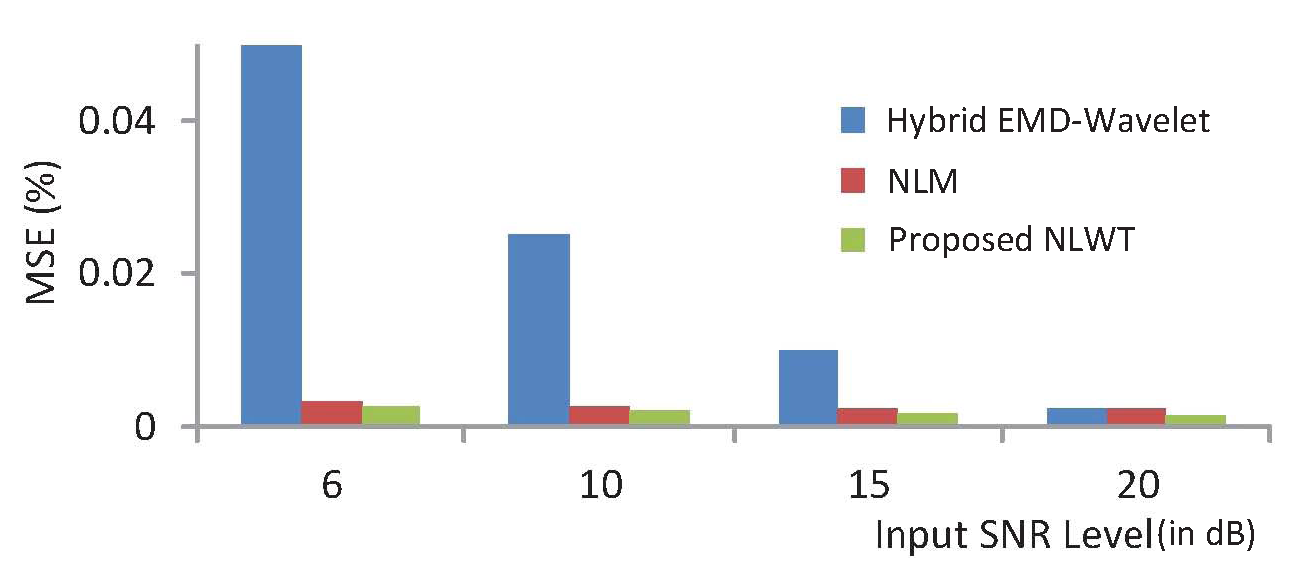}}
\subfigure{
\includegraphics[width=8cm,height=8cm]{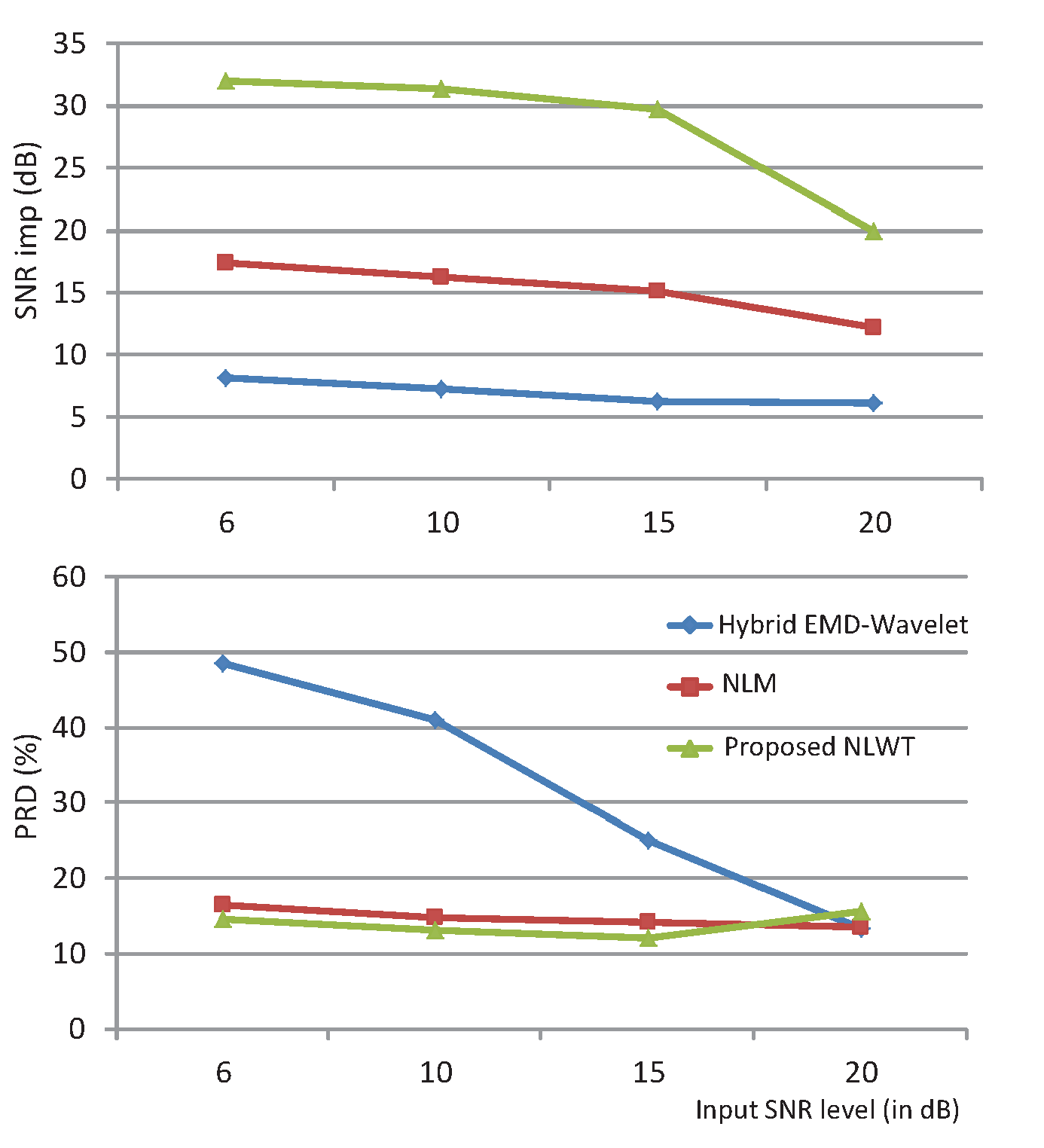}}
\end{center}
\caption{Comparison of the denoising performance of the proposed and the existing methods, in terms of different measures, averaged over the dataset at different SNR levels.}
\label{fig:perfecg3}
\end{figure}

\begin{figure*}[!ht]
\centering
\subfigure{
\includegraphics[width=18cm,height=9cm]{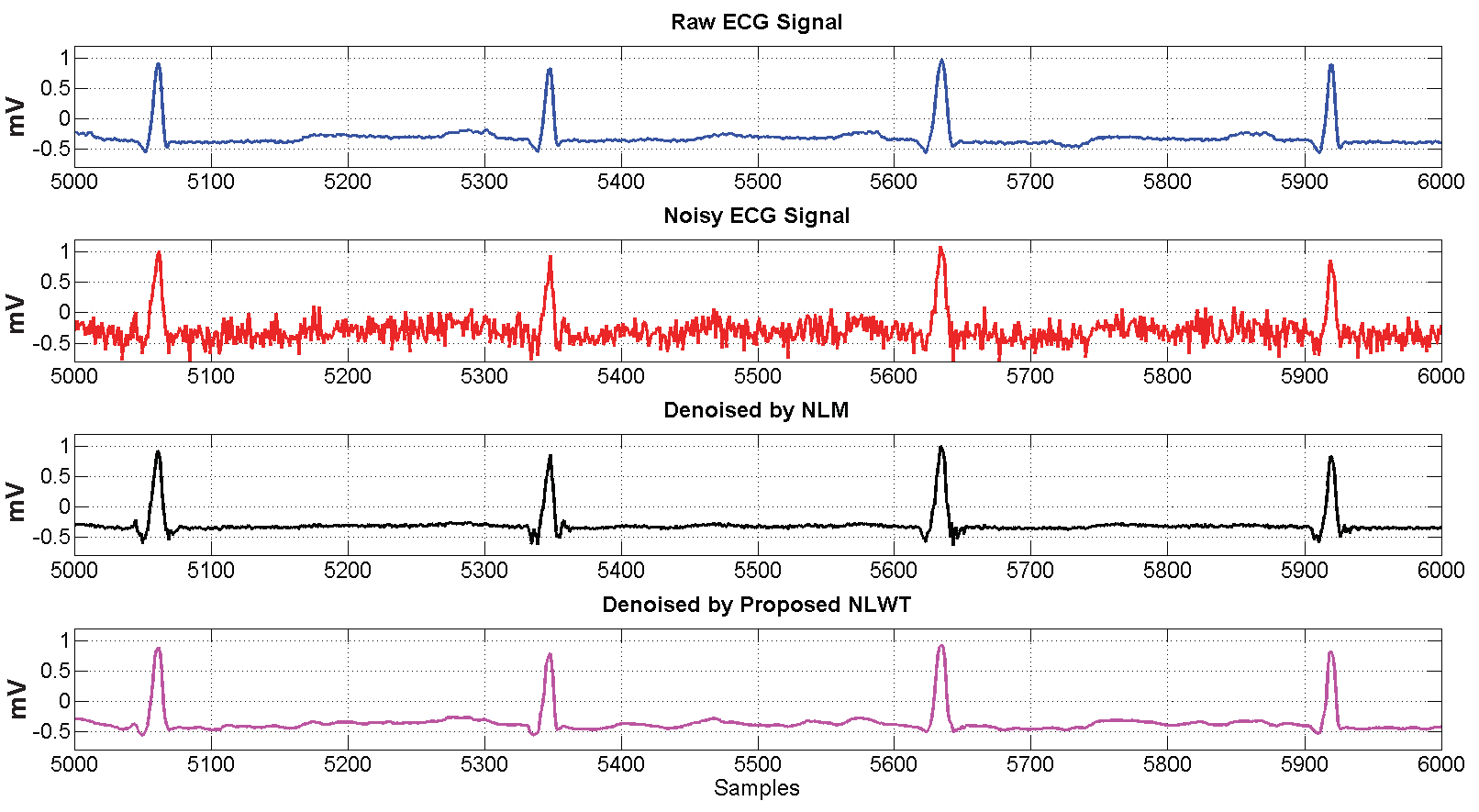}}
\caption{Waveforms showing a segment of the ECG signal number 100 for qualitative assessment. From top to bottom: raw (original) signal, 20 dB noisy signal, signal denoised using NLM algorithm, and signal denoised using proposed algorithm.}
\label{fig:perfecg1}
\end{figure*}

 For comparison, we have used the NLM~\cite{Tracey-Miller2012} and the hybrid wavelet-EMD~\cite{Kabir2012481} methods. These techniques are already shown to outperform the traditional wavelet thresholding and its variants in~\cite{Kabir2012481}. The NLM method is implemented using the parameter values given in \cite{Tracey-Miller2012} and the MATLAB code provided by its authors. For the hybrid-EMD method, the results are taken directly from \cite{Kabir2012481}. To be consistent with these works, five different realizations of the noise at four SNR levels (6 dB, 10 dB, 15 dB, and 20 dB), are added to the actual signals and the denoising performances averaged over these realizations are reported. The performances of the different methods on different signals, at 20 dB noise level, are given in Fig.~\ref{fig:perfecg2}. The performances averaged over the different signals of the chosen data set, at different noise levels are shown Fig.~\ref{fig:perfecg3}. Note that the NLWT method outperforms the existing methods in all three performance measures at all SNR levels. Among these measures, the improvement in SNR is considerably large for the proposed method.

 The biomedical signals contain a number of diagnostic features which have to be visually inspected by the medical practitioners. Hence, it is of prime importance to preserve those features. The important diagnostic features of an ECG signals are P-waves, QRS complexes, T-waves, U-waves, and their intervals. In denoising, these features usually undergo smoothing, thus affecting the diagnostic accuracy. Fig.~\ref{fig:perfecg1} shows the denoising performance of the NLM and the proposed NLWT method for a signal number 100 at 20 dB noise level. On comparing with the raw signal, it can be seen that the signal denoised by the NLWT method better resembles the clean signal, than the signal denoised by the NLM method.

\subsection{Evaluation on PTB diagnostic database}

\begin{figure}[!ht]
\centering
\subfigure{
\includegraphics[scale=0.4]{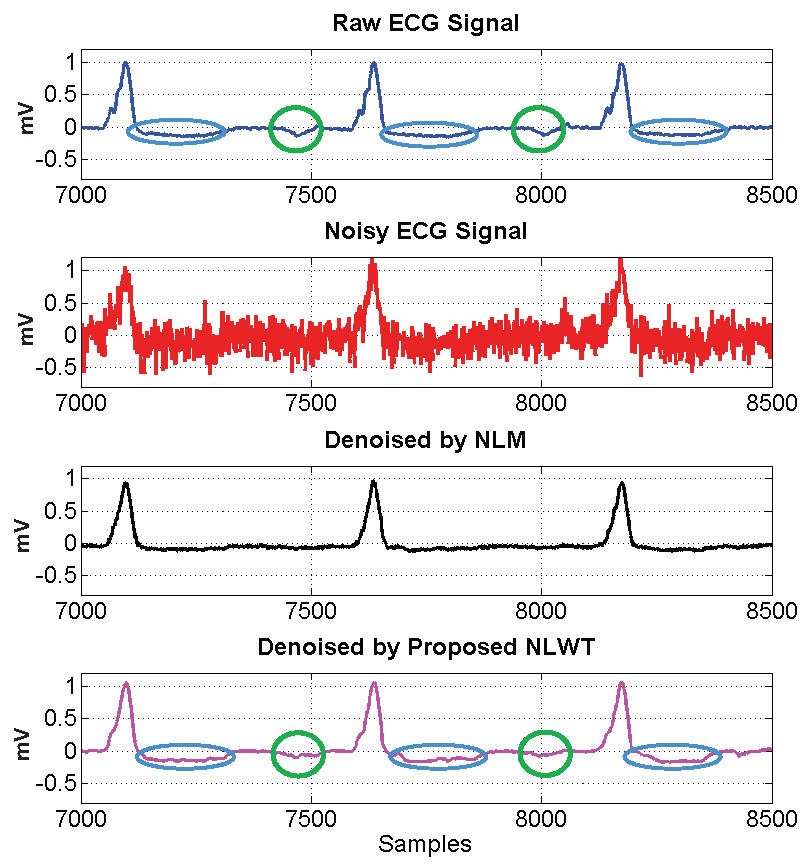}}
\caption{Illustration of preservation of the depressed ST-segments (marked with ellipse) and the inverted P-waves (marked with circle) in an ECG signal derived from record Id: s0370lrem of PTB diagnostic database. From top to bottom: raw (original) signal, noisy signal (30dB), signal denoised using NLM algorithm, and signal denoised using proposed algorithm. Note that unlike the NLM method, the marked diagnostic features are significantly preserved in case of proposed NLWT method.}
\label{fig:perfecg4}
\end{figure}

\begin{figure}[!ht]
\centering
\subfigure{
\includegraphics[width=8.5cm,height=9.2cm]{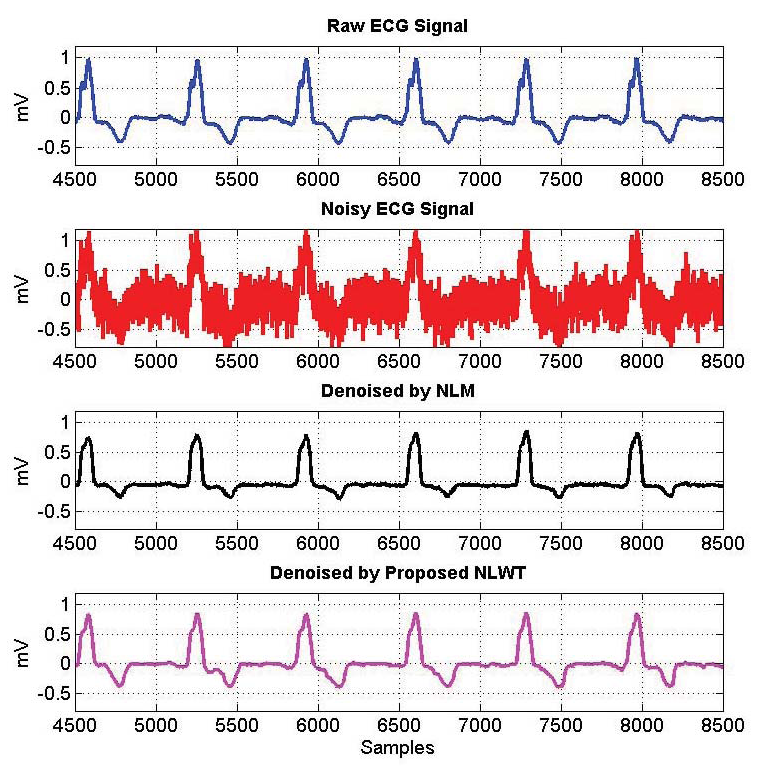}}
\caption{Illustration of inverted T-wave preservation in an ECG signal derived from record Id: s0361lrem of PTB diagnostic database. From top to bottom: raw (original) signal, noisy signal (20dB), signal denoised using NLM algorithm, and signal denoised using proposed algorithm. Note that the inverted T-waves are severely smoothed in case of NLM method while well preserved in case of the NLWT method.}
\label{fig:perfecg5}
\end{figure}

 The signals in the PTB diagnostic database are sampled at 1000 Hz. It contains a number of ECG signals with known pathology. Out of those, six diseases are chosen and for each disease two signals are considered for the performance evaluation. Each record in this database contains 15 simultaneously measured signals (12 conventional and 3 frank leads). To show the effectiveness of the proposed NLWT method on these ECG signals, denoising is performed on all these leads of the chosen signals.  As the sampling frequency of signals in this database differs from that in Physionet database, the parameters are accordingly adjusted. The block half-width $L=20$, the search window half-width $M=4000$, and matching threshold $\tau=1.8$ is found to be optimal for this data set. Parameters for the NLM method is also optimized in a similar way for fair comparison. For the different ECG signals with known pathology and corrupted with 20 dB noise, the denoising performances are given in Table~\ref{table-tab2}. These performance are averaged over all the 15 leads of the records. Note that, the NLWT method has consistently outperformed the NLM method for all the pathological ECG signals considered. For qualitative evaluation, the denoising on a few signals from the chosen data set are shown in Fig.~\ref{fig:perfecg4} and Fig.~\ref{fig:perfecg5}. Note that in Fig.\ref{fig:perfecg4}, the proposed method preserved the depressed ST-segments and inverted P-waves better than the NLM method. Fig.~\ref{fig:perfecg5} shows the excellent preservation of inverted T-wave by the proposed method with minor local variations in the denoised signal.

\begin{table*}[!ht]
\caption{Comparison of denoising performance of proposed NLWT method with NLM method on different diseased ECG signals derived from PTB diagnostic database. Best results are in bold.}
\begin{center}
\renewcommand{\arraystretch}{1.2}
\begin{tabular}{|c|c||c|c|c||c|c|c|}
\hline
    \multirow{2}{*}{Cardiac disease} & \multirow{2}{*}{Record Id} &  \multicolumn{3}{c||}{NLM} &  \multicolumn{3}{c|}{Proposed NLWT}  \\ \cline{3-8}
       & & PRD $(\%)$ & MSE & \text{SNR}$_{imp}$ (dB) & PRD $(\%)$  & MSE & \text{SNR}$_{imp}$ (dB) \\ \hline

      \multirow{2}{*}{Cardiac dysrhythmia} & s0032\_rem & 23.50& 0.0015&17.38 &\textbf{19.10} &\textbf{0.0010} &\textbf{19.18} \\ \cline{2-8}
      & s0207\_rem &27.07& 0.0010& 15.24& \textbf{20.45} &\textbf{0.0005} & \textbf{17.68} \\ \hline

      \multirow{2}{*}{Myocarditis} & s0508\_rem & 28.14 &0.0015 &15.43 &\textbf{18.82} &\textbf{0.0006} &\textbf{18.92} \\ \cline{2-8}
      & s0510\_rem & 24.04& 0.0011& 17.08& \textbf{16.43} &\textbf{0.0005} & \textbf{20.38} \\ \hline

      \multirow{2}{*}{Bundle branch block} & s0430\_rem & 20.32& 0.0011& 18.38& \textbf{16.81} & \textbf{0.0007} & \textbf{20.03} \\ \cline{2-8}
      & s0035\_rem & 30.25 & 0.0028 & 14.70 & \textbf{20.67} & \textbf{0.0013} & \textbf{18.01} \\ \hline

      \multirow{2}{*}{Myocardial infraction} & s0354lrem & 24.75 & 0.0011 & 16.38 & \textbf{20.20} & \textbf{0.0007} & \textbf{18.15} \\ \cline{2-8}
      & s0370lrem & 26.62 & 0.0017 & 16.17 & \textbf{20.32} & \textbf{0.0010} & \textbf{18.51} \\ \hline

      \multirow{2}{*}{Valvular heart disease} & s0003\_rem & 24.75 & 0.0013 & 16.47 & \textbf{17.51} & \textbf{0.0006} & \textbf{19.48}\\ \cline{2-8}
      & s0012\_rem & 25.60 & 0.0015 & 16.25 & \textbf{16.60} & \textbf{0.0006} & \textbf{20.01} \\ \hline

      \multirow{2}{*}{Ventricular hypertrophy} & s0432\_rem & 20.55& 0.0008& 18.18 & \textbf{17.05} & \textbf{0.0006} & \textbf{19.81} \\ \cline{2-8}
      & s0390lrem & 27.95 & 0.0016 & 15.44 & \textbf{21.60} & \textbf{0.0009} & \textbf{17.67} \\ \hline
\end{tabular}
\label{table-tab2}
\end{center}
\end{table*}

%%%%%%%%%%%%%%%%%%%%%%%%%%%%%%%%%%%%%%%%%%%%%%%%%%%%%%%%
\section{Discussion}
\label{sec:Discussion}

 The proposed NLWT method is based on the reasonable assumption that the underlying clean signal has several similar blocks of samples. The blocks of samples with smaller amplitudes obtain more similar blocks than the higher ones, due to the lower dynamic range and the nature of ECG signals. The averaging is not able to remove the noise from the higher amplitude regions and hence NLM method is found to suffer from the \emph{rare patch} effect. The same trend is also noticed in NLWT method, but it is less prominent because denoising is done in the 2D DWT domain. Moreover, the \emph{rare patch} effect can be overcome by adopting the suggestions given in~\cite{Tracey-Miller2012} and the references therein. Further, the NLWT method uses weighted averaging of the multiple estimates where weights are proportional to the confidence in the estimation of that block. If the simple averaging were used in the aggregation step, the use of adaptive thresholding in the coefficient shrinkage step would have become necessary. However, the adaptive thresholding would have increased the overall complexity many fold and so it is not explored in this work. Algorithms are implemented in MATLAB 7.14 on a 64-bit 3.20 GHz Intel (R) core (TM) i5-3470 computer with 4 Gb RAM. For a signal length of 10,000 samples, the proposed NLWT method takes 10 seconds while the NLM implementation on same platform takes 15 seconds. This smaller the run-time in NLWT is attributed to the use of block based processing rather than sample based procesing. The NLM implementation uses a fast NLM algorithm proposed in~\cite{fastNLMDarbon2008}, hence further reduction in processing time can be achieved by developing similar optimized codes for NLWT method.

 The proposed NLWT method is a combination of the wavelet coefficients shrinkage based denoising method and the NLM algorithm. The similar combination has also been used for image denoising~\cite{BM3D2007} in which denoising is done in two steps. In the first step using the similar combination as in NLWT, an intermediate estimate of the clean signal is obtained to get an estimate of the SNR. Then in the second step, denoising is performed by empirical Wiener filtering. In the proposed NLWT method, no such two stage procedure is used and hence it is much simpler as well as different from the method proposed in~\cite{BM3D2007}.

 In the NLWT method, improved performance is achieved for the white Gaussian noise. However, at the time of recording the ECG signals are mostly corrupted by non-Gaussian or structured noise caused by motion or muscle activity and other external interferences. To remove such noises, several methods have been proposed. But in case of ambulatory tele-monitoring, ECG signals are likely to be corrupted by channel noise which is commonly modeled as the Gaussian noise. Hence, the proposed algorithm finds a relevant scope of application. For addressing different kinds of noises, the proposed NLWT method can be combined with the interference rejection methods as done in~\cite{BlancoVelasco2008}.

%%%%%%%%%%%%%%%%%%%%%%%%%%%%%%%%%%%%%%%%%%%%%%%%%%%%%%%%

\section{Conclusion}
\label{sec:conclusion}

 This paper proposed a novel NLWT method for ECG signal denoising by exploiting the local and nonlocal redundancy present in the signal. On comparing with the two recent methods, the hybrid-EMD \cite{Kabir2012481} and the NLM method \cite{Tracey-Miller2012}, the proposed method is found to give better performance in terms of a number of performance measures. It is also highlighted that the proposed algorithm is able to preserve the diagnostic features in the signal much better than the existing algorithms. Recently, the learned basis transforms are reported to represent the signals more sparsely. Hence, one possible extension of this work is the use of learned basis transforms instead of the DWT. The other extension could be the automatic selection of the different parameters used in the proposed method.
%%%%%%%%%%%%%%%%%%%%%%%%%%%%%%%%%%%%%%%%%%%%%%%%%%%%%%%%%%%%%%

\bibliographystyle{IEEEtran}
\bibliography{reference_biomedical}

\begin{thebibliography}{10}
\providecommand{\url}[1]{#1}
\csname url@samestyle\endcsname
\providecommand{\newblock}{\relax}
\providecommand{\bibinfo}[2]{#2}
\providecommand{\BIBentrySTDinterwordspacing}{\spaceskip=0pt\relax}
\providecommand{\BIBentryALTinterwordstretchfactor}{4}
\providecommand{\BIBentryALTinterwordspacing}{\spaceskip=\fontdimen2\font plus
\BIBentryALTinterwordstretchfactor\fontdimen3\font minus
  \fontdimen4\font\relax}
\providecommand{\BIBforeignlanguage}[2]{{%
\expandafter\ifx\csname l@#1\endcsname\relax
\typeout{** WARNING: IEEEtran.bst: No hyphenation pattern has been}%
\typeout{** loaded for the language `#1'. Using the pattern for}%
\typeout{** the default language instead.}%
\else
\language=\csname l@#1\endcsname
\fi
#2}}
\providecommand{\BIBdecl}{\relax}
\BIBdecl

\bibitem{Jing-Jhang1999}
J.~Bai, Y.~Zhang, D.~Shen, L.~Wen, C.~Ding, Z.~Cui, F.~Tian, B.~Yu, B.~Dai, and
  J.~Zhang, ``A portable {ECG} and blood pressure telemonitoring system,''
  \emph{Engineering in Medicine and Biology Magazine, IEEE}, vol.~18, no.~4,
  pp. 63--70, 1999.

\bibitem{Alesanco-Garcia2010}
A.~Alesanco and J.~Garc$\Acute{i}$a, ``Clinical assessment of wireless {ECG}
  transmission in real-time cardiac telemonitoring,'' \emph{IEEE Trans.
  Information Technology in Biomedicine}, vol.~14, no.~5, pp. 1144--1152, 2010.

\bibitem{RafiAhmed}
M.~Rahman, R.~A. Shaik, and D.~V. R.~K. Reddy, ``Efficient and simplified
  adaptive noise cancelers for {ECG} sensor based remote health monitoring,''
  \emph{IEEE Sensors Journal}, vol.~12, no.~3, pp. 566--573, 2012.

\bibitem{huang1998empirical}
N.~E. Huang, Z.~Shen, S.~R. Long, M.~C. Wu, H.~H. Shih, Q.~Zheng, N.-C. Yen,
  C.~C. Tung, and H.~H. Liu, ``The empirical mode decomposition and the hilbert
  spectrum for nonlinear and non-stationary time series analysis,'' \emph{Proc.
  Royal Soc. London. Series A: Math., Phy. Engg. Sci.}, vol. 454, no. 1971, pp.
  903--995, 1998.

\bibitem{Agante1999}
P.~M. Agante and J.~de~Sa, ``E{CG} noise filtering using wavelets with
  soft-thresholding methods,'' in \emph{Computers in Cardiology}, 1999, pp.
  535--538.

\bibitem{RSemani2007}
R.~Sameni, M.-B. Shamsollahi, C.~Jutten, and G.~Clifford, ``A nonlinear
  bayesian filtering framework for {ECG} denoising,'' \emph{IEEE Trans. Biomed.
  Engg.}, vol.~54, no.~12, pp. 2172--2185, 2007.

\bibitem{OSayadi2008}
O.~Sayadi and M.-B. Shamsollahi, ``{ECG} denoising and compression using a
  modified extended kalman filter structure,'' \emph{IEEE Trans. Biomed.
  Engg.}, vol.~55, no.~9, pp. 2240--2248, 2008.

\bibitem{BlancoVelasco2008}
M.~Blanco-Velasco, B.~Weng, and K.~E. Barner, ``E{CG} signal denoising and
  baseline wander correction based on the empirical mode decomposition,''
  \emph{Computers in Biology and Medicine}, vol.~38, no.~1, pp. 1--13, 2008.

\bibitem{Gao-Sultan-Hu-Tung2010}
J.~Gao, H.~Sultan, J.~Hu, and W.~wen Tung, ``Denoising nonlinear time series by
  adaptive filtering and wavelet shrinkage: A comparison,'' \emph{IEEE Signal
  Processing Letters}, vol.~17, no.~3, pp. 237--240, 2010.

\bibitem{Kestler1998}
H.~Kestler, M.~Haschka, W.~Kratz, F.~Schwenker, G.~Palm, V.~Hombach, and
  M.~Hoher, ``Denoising of high-resolution {ECG} signals by combining the
  discrete wavelet transform with the wiener filter,'' in \emph{Computers in
  Cardiology}, 1998, pp. 233--236.

\bibitem{Weng2006}
B.~Weng, M.~Blanco-Velasco, and K.~Barner, ``{ECG} denoising based on the
  empirical mode decomposition,'' in \emph{IEEE Int. Conf. of Engg. in Med. and
  Bio. Soc.}, Aug 2006, pp. 1--4.

\bibitem{Kabir2012481}
M.~A. Kabir and C.~Shahnaz, ``Denoising of {ECG} signals based on noise
  reduction algorithms in {EMD} and wavelet domains,'' \emph{Biomed. Signal
  Process. Control}, vol.~7, no.~5, pp. 481--489, 2012.

\bibitem{Buades2005}
A.~Buades, B.~Coll, and J.~M. Morel, ``A non-local algorithm for image
  denoising,'' in \emph{IEEE Conference on Computer Vision and Pattern
  Recognition}, vol.~2, 2005.

\bibitem{Tracey-Miller2012}
B.~Tracey and E.~Miller, ``Nonlocal means denoising of {ECG} signals,''
  \emph{IEEE Trans. Biomed. Engg.}, vol.~59, no.~9, pp. 2383--2386, 2012.

\bibitem{BM3D2007}
K.~Dabov, A.~Foi, V.~Katkovnik, and K.~Egiazarian, ``Image denoising by sparse
  3{D} transform-domain collaborative filtering,'' \emph{IEEE Trans. Image
  Process.}, vol.~16, no.~8, pp. 2080--2095, 2007.

\bibitem{Priyam2009}
P.~Chatterjee and P.~Milanfar, ``Clustering-based denoising with locally
  learned dictionaries,'' \emph{IEEE Trans. Image Process.}, vol.~18, no.~7,
  pp. 1438--1451, 2009.

\bibitem{Rajwade-Banerjee}
A.~Rajwade, A.~Rangarajan, and A.~Banerjee, ``Image denoising using the higher
  order singular value decomposition,'' \emph{IEEE Trans. Pattern Anal. Mach.
  Intell.}, vol.~35, no.~4, pp. 849--862, 2013.

\bibitem{Van-Kocher2009}
D.~Van De~Ville and M.~Kocher, ``Sure-based non-local means,'' \emph{IEEE
  Signal Processing Letters}, vol.~16, no.~11, pp. 973--976, 2009.

\bibitem{Zhang-Feng-Wang2013}
X.~Zhang, X.~Feng, and W.~Wang, ``Two-direction nonlocal model for image
  denoising,'' \emph{IEEE Trans. Image Process.}, vol.~22, no.~1, pp. 408--412,
  2013.

\bibitem{Donoho1995}
D.~Donoho, ``De-noising by soft-thresholding,'' \emph{IEEE Trans. Inf. Theory},
  vol.~41, no.~3, pp. 613--627, 1995.

\bibitem{DONOHOjohnsten}
D.~L. Donoho and J.~M. Johnstone, ``Ideal spatial adaptation by wavelet
  shrinkage,'' \emph{Biometrika}, vol.~81, no.~3, pp. 425--455, 1994.

\bibitem{MoodyMIT-BIHArrhy}
G.~Moody and R.~Mark, ``The impact of the {MIT-BIH} arrhythmia database,''
  \emph{Engineering in Medicine and Biology Magazine, IEEE}, vol.~20, no.~3,
  pp. 45--50, May 2001.

\bibitem{PTBdataset}
R.~Bousseljot, D.~Kreiseler, and A.~Schnabel, ``Nutzung der
  {EKG}-signaldatenbank {CARDIODAT} der {PTB} über das internet.''
  \emph{Biomedizinische Technik. (Biomed. Tech.)}, vol.~40, no.~1, pp.
  317--318, 1995.

\bibitem{PhysionetGlobal}
A.~Goldberger, L.~Amaral, L.~Glass, J.~Hausdorff, P.~Ivanov, R.~Mark,
  J.~Mietus, G.~Moody, C.-K. Peng, and H.~Stanley, ``Physiobank, physiotoolkit,
  and physionet: Components of a new research resource for complex physiologic
  signals.'' \emph{Circulation}, vol. 101, no.~23, pp. 215--220, June 2000.

\bibitem{fastNLMDarbon2008}
J.~Darbon, A.~Cunha, T.~Chan, S.~Osher, and G.~Jensen, ``Fast nonlocal
  filtering applied to electron cryomicroscopy,'' in \emph{5th IEEE
  International Symposium on Biomedical Imaging}, May 2008, pp. 1331--1334.

\end{thebibliography}
\end{document}